\documentstyle[11pt,paspconf]{article}

\begin{document}

\title{A Serendipitous Search for Hy-Redshift Ly$\alpha$ Emission: \\
 A Case Study of Two Sources at $z \simeq 3$}
\author{Curtis Manning, Daniel Stern, Hyron Spinrad \& Andrew J. Bunker}
\affil{Astronomy Department, University of California,
        Berkeley, CA  94720 {\tt email: cmanning@astro.berkeley.edu}}

\begin{abstract}

In the course of our on-going search for serendipitous high-redshift
Ly$\alpha$ emission in deep archival Keck spectra, we discovered two
Ly$\alpha$ emission line candidates in a moderate dispersion ($
\lambda/{\Delta \lambda} \simeq 1200$) spectrogram.  Both lines have
high equivalent width ($W_{\lambda}^{\rm obs} \geq 450 {\rm \AA}$),
low velocity dispersions ($\sigma_v\, \sim 60 ~ {\rm km\, s}^{-1}$),
and deconvolved effective radii $r_e\approx 1.0 h_{50}^{-1}$ kpc.
Their sizes and luminosities are suggestive of the primeval galaxy
model of Lin \& Murray (1992), based on the self-similar collapse of
an isothermal sphere.  We argue that the line emission is Ly$\alpha$,
and it is stellar in origin.  The sources are consistent with being
primeval.

\end{abstract}



\section{Introduction}

Strong Ly$\alpha$ line emission has been proposed as a possible
signature of primeval galaxies.  Though early modeling conjectured
that these emission lines would be highly luminous but diffuse,
searches on 4m-class telescopes found no such emission lines (Pritchet
1994; Thompson \& Djorgovski 1995).  Deep spectra at the Keck
telescopes, however, regularly reveal serendipitous high-equivalent
width, isolated emission lines (e.g., Stern et al. 1999; Hu et
al. 1998).  Indeed, the first confirmed galaxy at $z > 5$ was
discovered in this manner (Dey et al. 1998).  We report on two sources
with strong line emission identified on a single slitlet from deep
slitmask observations of the SSA22 field, with equivalent widths
$W_{\lambda}^{\rm obs} \geq 375 {\rm \AA}$.  The source targeted by
this slitlet is a color-selected Lyman-break galaxy (LBG); we analyze
it in parallel for comparison. We adopt $(\Omega,\Lambda,{\rm H_0}) =
(1,0,{50})$.

\section{Observations and Results}

We have obtained deep, moderate-dispersion spectra of $z \simeq 3$
LBGs in the SSA22 field with the aim of detailed studies of the ages,
kinematics, dust-content, and abundances of the LBG population (Dey et
al., in preparation).  The data were taken with the Low Resolution
Imaging Spectrometer (Oke et al. 1995) at the Keck II telescope on UT
1997 September 10, using the 600 lines$~{\rm mm}^{-1}$ grating blazed
at $5000 {\rm \AA}$.  Slitlet widths were $1\farcs 25$, resulting in a
spectral resolution of $\sim 4.4{\rm \AA}$ (FWHM).  Four $1800~s$
exposures (seeing FWHM$\simeq 0 \farcs 78$) are analyzed in this
report.  The data were processed using standard slit spectroscopy
procedures (for details, see Manning et al., in preparation).

\begin{figure}[h]
  \plotfiddle{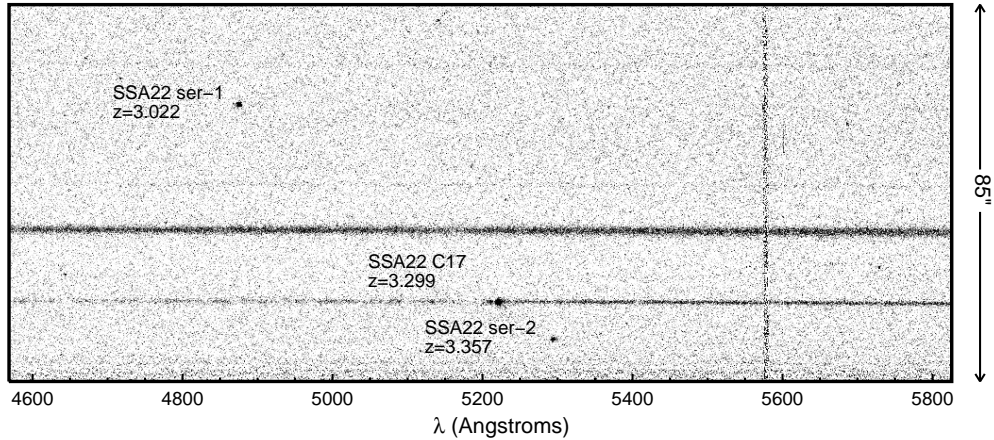}{2.0in}{0}{91}{91}{-190}{-21} 
\caption
 { Two-dimensional spectrogram of slitlet in the SSA22 field targeting
 the LBG C17 at z=3.299.  Two strong line emitters are serendipitously
 identified, ser-1 and ser-2.  The high-equivalent widths, narrow
 velocity widths, and lack of seconydary spectral features strongly
 argue that these are Ly$\alpha$ emitters at $z\simeq 3$.  Note the
 foreground continuum source and residual of the [O {\scriptsize
 I}]5577{\rm \AA} skyline.}
\end{figure}

We find two serendipitous, isolated emission lines, ser-1 and ser-2,
in a $100^{\prime\prime} $ long slitlet centered on the LBG, C17
(z=3.299; see Fig. 1).  Though only one of these serendipitously
discovered sources (see Fig. 2) displays the obvious asymmetry
characteristic of most high-redshift Ly$\alpha$ emission (e.g., Dey et
al. 1998), we argue that Ly$\alpha$ is indeed the most likely
interpretation for both.  We performed a simulation of the [O
{\scriptsize II}] doublet under the $4.4 {\rm \AA}$ resolution of our
observations.  We find that an [O {\scriptsize II}] identification is
inconsistent with ser-1 and ser-2, since the doublet is marginally
resolved and has significantly greater FWHM than our lines.  The
absence of associated emission argues against H${\beta}$, [O
{\scriptsize III}]4959, or [O {\scriptsize III}]5007 interpretations.
Finally, the lines are shortward of $6563 {\rm \AA}$, so H${\alpha}$
is not a viable identification.  In the following, we assume what is
most certainly the case --- that these are in fact Ly$\alpha$ emission
lines.  Notably, neither spectrum displays perceptible continuum.

Measured continua were consistent with zero counts.  We establish a $2
\sigma$ ($95 \%$ confidence) limit on the equivalent width,
$W_{\lambda}^{\rm obs} \geq 550 {\rm \AA}$ (ser-1), and $\geq 474 {\rm
\AA}$ (ser-2).  Rest values are tabulated in Table 1.  C17 has
$W_{\lambda}^{\rm obs} = 35.6 {\rm \AA}$.

\begin{figure}[h]
\plotfiddle{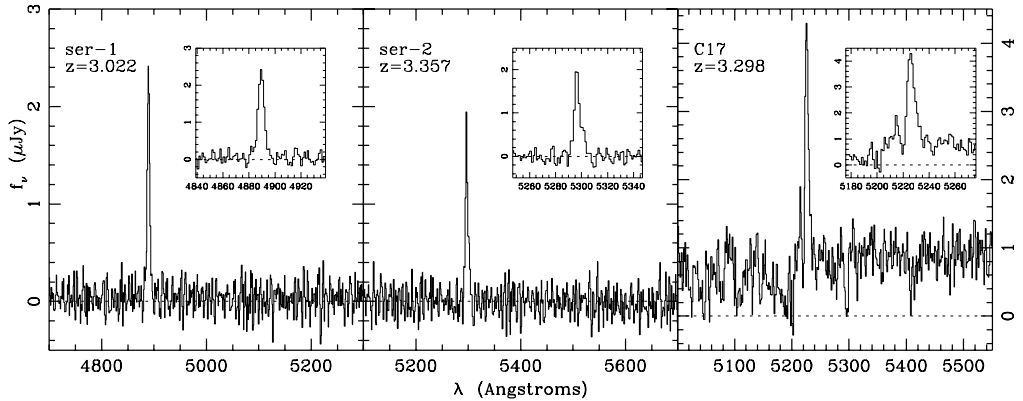}{2.0in}{0}{70}{70}{-215}{-340} 
\caption{ Extracted spectra of ser-1, ser-2 and the targeted LBG, C17.
The ordinate on the left refers to the serendipitous sources, while
the ordinate on the right refers to C17.  Inserts illustrate the
characteristic asymmetric Ly$\alpha$ profile in ser-2 and C17.}
\end{figure}

\begin{table}
\caption{Emission Line Properties} 
\label{tbl-1}
\begin{center}\scriptsize
\begin{tabular}{cccccccccc}
\hline\hline {Object} & {$\lambda^{\rm obs}$ } & {z} & {$j_{-17}$} &
{$L_{42}$} & ${W_{\lambda}^{\rm obs}}$ & {${\rm
 W_{\lambda}^{\rm rest}}$} & {$r_e$} & {$\sigma_v$} \\
- & $({\rm \AA})$ & - & ${\rm cgs}$ & ${\rm cgs}$ &
 $({\rm \AA})$ & 
$({\rm \AA})$ & $(h_{50}^{-1}$ kpc) & $({\rm km \, s^{-1}})$ \\
\tableline
ser-1 &4889.3 &3.022 &1.85 &1.29 &$\geq 550$ &$\geq 137$ &0.87
 &64 \\ 
ser-2 &5296.8 &3.357 &1.35 &1.19  &$\geq
 474 $ &$\geq 109$ &1.15 &52 \\ 
C17 &5226.4 &3.299 &2.60 &2.20 &35.6 &8.3 &2.79 &
 82 \\
\hline
\end{tabular}
\end{center}

\end{table}

Intrinsic source sizes are approximated by deconvolution of the
observed spatial FWHM with the seeing, obtained from a quasar in an
adjacent slitlet.  The velocity widths have been deconvolved using the
instrumental resolution, obtained from spectra of a NeAr lamp (see
Table 1).  The velocity dispersions of ser-2 and C17 are lower
limits, as possible truncation of the blue-side of the emission line
has not been accounted for.

\section{Discussion and Conclusions}

What is the physical origin of these isolated, high-equivalent width
emission lines?  The small velocity dispersions ($\sigma_v \simeq
60~{\rm km ~s}^{-1}$) and lack of associated N~{\scriptsize
V} or C~{\scriptsize IV} emission argue against the presence of an
AGN, while the high surface brightness is inconsistent with
photoionization by the metagalactic flux (e.g., Bunker et al. 1998).
Thus we assume the lines are stellar in origin.  Below we argue that
they are primeval.

The starburst modeling of Charlot \& Fall (1993) suggest that the
Ly$\alpha$ rest equivalent width may exceed $\sim 150 {\rm \AA}$ for
timescales of at most a few $\times 10^7$ yr.  A model of primeval
galaxies based on a collapsing isothermal cloud (Lin \& Murray 1992)
predicts that primeval galaxies should have isophotal radii $r_e \sim
1~ {\rm kpc}$, and luminosities of order $\sim {\cal L}^*$; strong
star formation continues for $\sim 1.7~{\rm Gyr}$, fed by the
continuing collapse of the cloud.  Cloud mass may be tuned to fit
observed luminosities.  We suggest that this theory holds promise for
the modeling of our serendipitous sources.

The Ly$\alpha$ luminosities imply lower limits to star formation rate
(SFR) of $0.87/0.81~{\rm M_{\odot} ~yr^{-1}}$ for ser-1/ser-2 (e.g.,
Dey et al. 1998), while non-detection of continua imply upper limits
of $2.6/2.4~{\rm M_{\odot} ~yr^{-1}}$ .  Assuming a SFR of
$1~{M_\odot}{\rm yr}^{-1}$, and $W_{\lambda}^{\rm rest}=150 {\rm
\AA}$, we project $V_{\rm AB} \simeq 28.0$ at $z = 3$.  Steidel et
al. (1999) find a steep faint end slope for the luminosity function of
star-forming galaxies at $z \sim 3$ ($\alpha \approx -1.6$), implying
that a large fraction of the UV luminosity density is produced by
galaxies fainter than the spectroscopic limits of photometric surveys.
By virtue of the association of UV luminosity with SFR, this would
also imply that a large fraction of the star formation density is
unobserved.  Because the Ly$\alpha$ lines of such as our serendipitous
objects are likely to be observable only for a very small fraction of
1 Gyr, these two are tangible evidence for this many-fold larger
population of slightly older, faint continuum, star forming galaxies
with self-absorbed Ly$\alpha$, inaccessible to even the deepest
spectrosopic or color selection surveys.

The evidence strongly suggests that ser-1 and ser-2 are primeval.  In
the near future, interesting conclusions may be possible regarding the
degree of clustering and/or isolation of these low-luminosity emission
line galaxies.  The large surface density variance in Ly$\alpha$
emission line galaxy candidates between different fields in a coarse
``narrow-band'' ($z \simeq 2.37$) HST study (Pascarelle et al. 1998),
indicates that these compact galaxies are often at least loosly
grouped.  Clearly, the model of Lin \& Murray requires that primeval
galaxies be born in relatively quiescent, yet gas-rich environments.
However, as star-forming galaxies, their likely evolutionary
antecedents, the LBGs, are often found clustered on large scales
(e.g., Steidel et al. 1998); perhaps they are protoclusters.  The
dynamics involved in the growth of large scale perturbations (i.e.,
infall and rising central densities) may be accompanied by a spreading
``wave'' (in the Lagrangian sense) of primeval galaxy formation,
perhaps occurring at a characteristic overdensity --- a scenario
reminiscent of stellar ``shell burning'' --- reproducing the inside-out,
older-to-younger galaxy age distribution seen in some recent studies
of lower redshift fields and groups (e.g., Loveday et al. 1999; Cohen
et al. 1998).

\acknowledgments We thank C. C. Steidel and A. Dey for useful
discussions and assistance with the observations.

\end{document}